\documentclass[twocolumn,showpacs,preprintnumbers,amsmath,amssymb]{revtex4}
\usepackage{graphicx}
\usepackage{dcolumn}
\usepackage{bm}
\begin{document}
\title{Fluctuation-dissipation theorems for viscoelastic fluids:\\
consistency of extended thermodynamic theories}
\author{F. V{\'a}zquez}%
 \email{vazquez@servm.fc.uaem.mx}
\author{M.A. Olivares-Robles}%
 \email{olivares@servm.fc.uaem.mx}
\author{S. Delgado}
\affiliation{Departamento de F{\'\i}sica, Facultad de Ciencias,
Universidad Aut\'onoma del Estado de Morelos, Av. Universidad 1001,
Col. Chamilpa, Cuernavaca, Morelos 62210, M\'exico}%

\date{\today}

\begin{abstract}
Fluctuation-Dissipation Relations (FDR) for a Maxwell fluid are computed via the GENERIC formalism. This formalism is determined by four building blocks, two ``potentials'' (total energy and entropy) and two ``matrices'' which determine the dynamics, but the understanding of fluctuations in a given non-equilibrium system is reduced to determining a single friction matrix. The FDR exhibits interesting features, arising from the type of entropy used in the formalism. We show explicitly this dependence FDR-Entropy, also we show that GENERIC renders results consistent with irreversible linear thermodynamics (TIL) \emph{only} if it is used the corresponding entropy. An inconsistent result is provided by GENERIC when it is used a modified entropy .
\end{abstract}
\pacs{Valid PACS appear here}
\maketitle

\section{INTRODUCTION}
Much effort has been expended within the theory of stochastic processes to obtain the mesoscopic basis of macroscopic theories. The problem of the description of equilibrium and non-equilibrium fluctuations in macroscopic systems is one of the central aspects of both thermodynamic and microscopic theories of irreversible processes \cite{Lax60}. From the macroscopic viewpoint, the non-equilibrium case has been investigated within the framework of several theories among which we mention in particular extended irreversible thermodynamic theory (EIT) \cite{Jou96}. The starting point of extended thermodynamics is the generalization of the Gibbs relation for the non-equilibrium entropy, which is used to determine the second moments of the physical fields under the assumption that the probability of the fluctuations is given by the Einstein relation \cite{Joujnet80}. Another, general formalism for studying the dynamics of non-equilibrium systems is called GENERIC (general equation for the non-equilibrium reversible irreversible coupling) \cite{generic}. This framework has been developed to formulate dynamic equations for out-of-equilibrium systems, by empirical arguments and by identifying the common structure of successful models in non-equilibrium thermodynamics and it has been validated by projection operator techniques \cite{Ottinger1}. In particular, Generic has proved to be a strong framework for dealing with the modeling of rheological properties of complex fluids \cite{Ottinger02}.\\
Recently, in an attempt to test the consistency between alternative theories of irreversible processes, the equilibrium Rayleigh-Brillouin spectrum for a Maxwell fluid has been computed. This was made by using fluctuating hydrodynamics, via fluctuation-dissipation relations (FDR), and by the procedure used by Mountain \cite{Mariano01} which is a direct calculation technique. One would a priori have expected that since the Maxwell model has been often quoted as the prototype system for an EIT treatment, the equivalence of the two routes would have occurred when performing both types of calculations. However, a discrepancy arose whose origin could be traced to the fluctuation-dissipation relation for the (stochastic) stress tensor. If the prescription made by Landau for the frequency-dependent dissipative coefficients case is used both routes coincide. In contrast, if the FDR from EIT is used the two routes diverge.\\
In this work, we undertake again the testing of consistency between alternative thermodynamic theories. We describe fluctuations in a viscoelastic fluid by using three different thermodynamic frameworks, namely, linear irreversible thermodynamics (LIT), and two versions of extended irreversible thermodynamics. The first one is extended with the stress tensor, and the second one with the conformation tensor of the kinetic theory of polymeric fluids \cite{generic}. Our purpose in this paper is to discuss explicitly, via GENERIC, why the LIT framework must to be used for calculating a FDR in rheological models for complex fluids instead of the extended versions of irreversible thermodynamics. An unavoidable conclusion is arising from this analysis. The extended versions of fluctuating irreversible thermodynamics must be revised.\\
The paper is organized as follows. Section II is devoted to describe the main characteristics of GENERIC to make this article self-contained. In Section III we illustrate how the FDR are obtained for a known case, namely, a Newtonian fluid which is described by LIT. We analyze in Section IV the case of a viscoelastic fluid within LIT and EIT by extending the thermodynamic variable space with the dissipative fluxes and with the conformation tensor \cite{generic}. We obtain the corresponding FDR for each case. Finally, in Section V we make some discussion on the previous issues and add some conclusions.\\
\section{FLUCTUATION-DISSIPATION RELATIONS AND GENERIC}
The fundamental time-evolution equation of GENERIC, which was developed by considering the compatibility of two levels of description and by studying a large number of specific examples \cite{generic}, can be written in the form
\begin{equation}
\frac{dx}{dt}=L(x)\frac{\delta E(x)}{\delta x}+M(x)\frac{\delta S(x)}{\delta x},\label{generic}%
\end{equation}
where $x$ represents the set of independent state variables required for a complete description of the underlying non-equilibrium system, the real-valued functionals $E$ and $S$ are the total energy and entropy expressed in terms of the state variables $x$, and $L$ and $M$ are the Poisson and friction matrices (or linear operators). The two contributions to the time-evolution of $x$ generated by the energy $E$ and the entropy $S$ in Eq. \ref{generic} are called the reversible and irreversible contributions to GENERIC, respectively. Since $x$ typically contains position-dependent fields, such as mass, momentum and energy densities, the state variables are usually labeled by continuous labels in addition to discrete ones. A matrix multiplication, or the application of a linear operator, hence implies not only summations over discrete labels but also integrations over continuous labels, and $\delta/\delta x$ typically implies functional rather than only partial derivatives.\\
In the GENERIC framework, Eq. (\ref{generic}) is supplemented by the complementary degeneracy requirements
\begin{equation}
L(x)\frac{\delta S(x)}{\delta x}=0,\label{degeneracion1}
\end{equation}
\begin{equation}
M(x)\frac{\delta E(x)}{\delta x}=0.\label{degeneracion2}
\end{equation}
The requirement (\ref{degeneracion1}) expresses the reversible nature of the $L$-contribution to the dynamics: the functional form of the entropy is such that it cannot be affected by the operator generating the reversible dynamics. The requirement (\ref{degeneracion2}) expresses the conservation of the total energy by the $M$-contribution to the dynamics. Furthermore, it is required that the matrix $L$ is antisymmetric, whereas $M$ is Onsager-Casimir symmetric and semi-positive definite. Both the complementary degeneracy requirements (\ref{degeneracion1}), (\ref{degeneracion2}) and the symmetry properties are extremely important for formulating proper $L$ and $M$-matrices when modeling concrete non-equilibrium problems \cite{Ottinger1}. Finally, the Poisson bracket associated with the antisymmetric matrix $L$,
\begin{equation}
\left\{A,B\right\}=\left<\frac{\delta A}{\delta x},\ L(x)\frac{\delta B}{\delta x}\right>,\label{parentesispoisson}
\end{equation}
with
\begin{equation}\label{bracket}
\{A,B\}=-\{B,A\}
\end{equation}
is assumed to satisfy the Jacobi identity,
\begin{equation}
\left\{\left\{A,B\right\},C\right\}+\left\{\left\{B,C\right\}
,A\right\}+\left\{\left\{C,A\right\},B\right\}=0,\label{jacobi}
\end{equation}
for arbitrary functionals A, B, and C. This identity, which expresses the time-structure invariance of the reversible dynamics, is postulated as another important general property required by non-equilibrium thermodynamics \cite{generic,Edwars97}.\\
The stochastic dynamics into GENERIC is determined by the stochastic differential equation obtained by adding noise (and the divergence of $M$) to Eq. (\ref{generic}) \cite{Ottinger96},
\begin{equation}
dx=L\frac{\delta E}{\delta x}+M\frac{\delta S}{\delta
x}+k_{B}\frac{\delta M}{\delta x}dt+B(x)dW_t,\label{langevin1}
\end{equation}
where $B$ is a solution of the equation
\begin{equation}
BB^{T}=2k_{B}M.\label{raiz}
\end{equation}
and $W_t$ is a multi-component Wiener process, that is, a Gaussian process with first and second moments given by
\begin{equation}
\left\langle\frac{dW_t}{dt}\right\rangle=0, \quad \left\langle\frac{dW_t}{dt}\frac{dW^{T}_{t'}}{dt'}\right\rangle=\delta(t-t')\mathbf{1}.
\end{equation}
The expression (\ref{raiz}) for $B$ may be regarded as the \emph{fluctuation-dissipation relation}.\\
We now show how the FDR may be obtained in this framework. Eq. (\ref{langevin1}) is written by components as follows
\begin{equation}
dx_{r}^{m}(t)=A_{r}^{m}(x)dt+B_{rr^{\prime}}^{mm^{\prime}}(x) dW_{r^{\prime}}^{m^{\prime}}(t)\label{componentes},%
\end{equation}
where the definition of $A_{r}^{m}(x)$ is evident and the increments $dW$ satisfy the property
\begin{equation}
dW_{r}^{m}(t)dW_{r^{\prime}}^{m^{\prime}}(t)=
\begin{cases}
\delta_{mm^{\prime}}\delta_{rr^{\prime}}dt&\text{if $t=t^{\prime}$}\\
0&\text{if $t\neq t^{\prime}$}\label{incrementos}
\end{cases}
\end{equation}
The Kronecker's delta $\delta_{rr^{\prime}}$ is equal to unity if
$r=r^{\prime}$ and vanishes otherwise. In Eq. (\ref{componentes}) a white noise term defined as
\begin{equation}
\eta^{m}(\mathbf{r},t)\equiv\frac{dW_{r}^{m}(t)}{dt}\label{blanco}
\end{equation}
may be introduced. This term has the following correlation function
\begin{equation}\label{correlacion}
\begin{split}
\left\langle \eta^{m}(\mathbf{r},t)\eta^{m^{\prime}}(\mathbf{r}^{\prime},t^{\prime})\right\rangle &\equiv\left\langle\frac{dW_{r}^{m}(t)}{dt}\frac{dW_{r^{\prime}}^{m^{\prime}}
(t^{\prime})}{dt^{\prime}}\right\rangle\\
&=
  \begin{cases}
    \delta_{mm^{\prime}}\delta_{rr^{\prime}}\frac{1}{dt}& \text{if $t=t^{\prime}$}, \\
    0& \text{if $t\neq t^{\prime}$}.
  \end{cases}
\end{split}
\end{equation}
This expression may be rewritten as
\begin{equation}
\left\langle \eta^{m}(\mathbf{r},t)\eta^{m^{\prime}}(\mathbf{r}^{\prime
},t^{\prime})\right\rangle =\delta_{mm^{\prime}}\delta(r-r^{\prime}%
)\delta(t-t^{\prime}).\label{delta}%
\end{equation}
\noindent which gives the same results as Eq. (\ref{correlacion}). We define the stochastic term in Eq. (\ref{componentes}) as
\begin{equation}
f_{r}^{m}(t)dt\equiv\sum_{r^{\prime}}B_{rr^{\prime}}^{mm^{\prime}%
}(x(t))dW_{r^{\prime}}^{m^{\prime}}(r,t).\label{ruidof}%
\end{equation}
Calculating its correlation we obtain
\begin{widetext}
\begin{equation}
\begin{split}
\left\langle \int dt\int dt^{\prime\prime}f_{r}^{m}(t)f_{r^{\prime\prime}%
}^{m^{\prime\prime}}(t^{\prime\prime})\right\rangle \label{correlacionf}\\
&=\left\langle \int\int\sum_{r^{\prime}}B_{rr^{\prime}}^{mm^{\prime}%
}(x(t))dW^{m^{\prime}}(r^{\prime},t)\times\sum_{r^{\prime\prime\prime}%
}B_{r^{\prime\prime}r^{^{\prime\prime\prime}}}^{m^{\prime\prime}%
m^{\prime\prime\prime}}(x(t^{\prime}))dW^{m^{\prime\prime\prime}}%
(r^{\prime\prime\prime},t^{\prime})\right\rangle\\
&=\left\langle \int dt\sum_{r^{\prime}}B_{rr^{\prime}}^{mm^{\prime}%
}(x(t))B_{r^{\prime\prime}r^{\prime}}^{m^{\prime\prime}m^{\prime}%
}(x(t))\right\rangle\\
&=2k_{B}\int dt\int Dxf(x,t)M_{rr^{\prime}}^{mm^{\prime\prime}}(\alpha
)=2k_{B}\int dt\left\langle M_{rr^{\prime}}^{mm^{\prime\prime}}\right\rangle
_{t},
\end{split}
\end{equation}
\end{widetext}
where we have made use of Eq. (\ref{incrementos}), $\ f(x,t)$ is the probability density that the system takes the values $\alpha$ at time $t$ \cite{espanol}$.$ Moreover, note that
\begin{equation}
\left\langle f_{r}^{m}(t)f_{r^{\prime\prime}}^{m^{\prime\prime}}%
(t^{\prime\prime})\right\rangle =2k_{B}\delta(t-t^{\prime\prime})\left\langle
M_{rr^{\prime\prime}}^{mm^{\prime\prime}}\right\rangle _{t}\label{deltacorrelacion}%
\end{equation}
\noindent since if we integrate on $t,t^{\prime\prime}$ we recover Eq. (\ref{correlacionf}) and the integration has arbitrary limits. Expression (\ref{deltacorrelacion}) is another way to write the FDR. We will use this expression in the following Sections. As it has been mentioned the form of matrix $M$ depends on the form of the entropy out from equilibrium. Therefore it determines the specific form adopted by expression (\ref{deltacorrelacion}).
\section{THE VISCOUS FLUID IN LIT}
As an illustration of the stochastic dynamics (\ref{langevin1}) and (\ref{raiz}), we consider an example of fluctuating hydrodynamics. The results of the previous Section are illustrated with the case of a viscous Newtonian fluid within LIT. As usual the physical conserved densities are chosen as independent variables to describe the thermodynamic state of the fluid. Mass density is denoted by $\rho$, momentum density by $\mathbf{u}$, and internal energy density by $\varepsilon.$ The balance equations of mass, momentum and energy are written as
\begin{equation}
\frac{\partial}{\partial t}\rho=-\frac{\partial}{\partial\mathbf{r}%
}\cdot(\mathbf{v}\cdot\rho)\label{masa}%
\end{equation}
\begin{equation}
\frac{\partial\mathbf{u}}{\partial t}=-\frac{\partial}{\partial\mathbf{r}%
}\cdot(\mathbf{v}\cdot\mathbf{u})-\frac{\partial}{\partial\mathbf{r}}p-\frac{\partial
}{\partial\mathbf{r}}\cdot\overleftrightarrow{\mathbf{\tau}}\label{momento}%
\end{equation}
\begin{equation}
\frac{\partial\mathbf{\varepsilon}}{\partial t}=-\frac{\partial}%
{\partial\mathbf{r}}\cdot(v\cdot\mathbf{\varepsilon})-p\frac{\partial}{\partial
\mathbf{r}}\cdot\mathbf{v}-\frac{\partial}{\partial\mathbf{r}}\cdot\mathbf{q}%
-\overleftrightarrow{\mathbf{\tau}}\mathbf{:}\frac{\partial}{\partial
\mathbf{r}}\mathbf{v}\label{energia}%
\end{equation}
\noindent where $p$ is pressure, the expression for the stress tensor is that of Newton and it is written in terms of the velocity gradient, shear viscosity $\eta$, and volume viscosity $\zeta$ as follows
\begin{equation}
\overleftrightarrow{\mathbf{\tau}}=-\eta\left[  \frac{\partial}{\partial
\mathbf{r}}\mathbf{v+}\left(  \frac{\partial}{\partial\mathbf{r}}
\mathbf{v}\right) ^{T}\right] -\left(  \zeta-\frac{2}{3}\eta\right)
\frac{\partial}{\partial\mathbf{r}}\cdot\mathbf{v}\overleftrightarrow{\mathbf{1}%
}\label{newtoniano}
\end{equation}
\noindent Fourier's heat conduction is assumed and written in terms of the temperature gradient and thermal conductivity $K$
\begin{equation}
\mathbf{q}=-K\frac{\partial T}{\partial\mathbf{r}}\label{fourier}
\end{equation}
Knowing the temperature $T=T(\rho,\varepsilon)$ and pressure
$p=(\rho,\varepsilon)$, Eqs. (\ref{masa}-\ref{fourier}) form a closed set of time evolution equations. The irreversible contributions to the dynamics are those involving the stress tensor $\overleftrightarrow{\mathbf{\tau}}$ and the heat flux $\mathbf{q}$ appearing in Eqs. (\ref{momento}) y (\ref{energia}); the remaining terms are the reversible contributions to the dynamics. The dissipative coefficients may depend on position. Though the matrix of the reversible part of the dynamics is not necessary for the study of fluctuations we include a brief summary to give the complete GENERIC's equation. The total energy needs to be expressed as a function of the set of variables of the system. This set is denoted by the vector $x=(\rho,\mathbf{u},\varepsilon)$. The vector $\delta E/\delta x$ results in a vector with five entries constituted by functional derivatives of the energy with respect to $x $. The total energy is obtained by integrating the addition of the kinetic energy and internal energy densities over the volume of the system
\begin{equation}
E=\int\left[  \frac{1}{2}\frac{\mathbf{u(r)}^{2}}{\rho}+\mathbf{\varepsilon
(r)}\right]  d^{3}r\label{energiatotal}%
\end{equation}
The functional derivative of this energy with respect to $x,$\ is given by%
\begin{equation}
\frac{\delta E}{\delta x}=
\left(
\begin{array}
[c]{c}
\frac{\delta}{\delta\rho(r)}\\
\frac{\delta}{\delta\mathbf{u}(r)}\\
\frac{\delta}{\delta\varepsilon(r)}
\end{array}
\right) E(\rho\mathbf{,u,}\varepsilon)=\left(
\begin{array}
[c]{c}
-\frac{1}{2}\mathbf{v(r)}^{2}\\
\mathbf{v(r)}\\
1
\end{array}
\right) \label{energiagradiente}%
\end{equation}
The matrix $L$ may be constructed in two ways. Both have been explored in detail in \cite{generic} and \cite{beris91}. We only mention here that the last authors have based their method on variational techniques. To determine the dissipation operator, the functional of entropy of LIT is considered
\begin{equation}
S=\int\mathbf{s}(\rho(\mathbf{r}),\mathbf{\varepsilon}(\mathbf{r}%
))d^{3}\mathbf{r}\label{entropiatil}%
\end{equation}
\noindent in such a way that the functional derivative with respect to the hydrodynamics fields is given by
\begin{equation}
\frac{\delta S}{\delta x}=\left(
\begin{array}
[c]{c}%
-\frac{\mu(\mathbf{r)}}{T(\mathbf{r})}\\
\mathbf{0}\\
\frac{1}{T(\mathbf{r})}%
\end{array}
\right) \label{entropiagradiente}%
\end{equation}
\noindent where the local temperature and the chemical potential are defined as
\begin{equation}
\left(\frac{\partial s(\rho\mathbf{,\varepsilon})}{\partial
\mathbf{\varepsilon}}\right)_{v}=1/T(\mathbf{r)}\label{definicion}
\end{equation}
\begin{equation}
\frac{\mu(\mathbf{r)}}{T(\mathbf{r)}}=-\frac{\partial\mathbf{s(}
\rho\mathbf{,\varepsilon)}}{\partial\rho}\label{definicion2}
\end{equation}
The matrix $M$,\ which reproduces the irreversible terms in the dynamic equations, then takes the form
\begin{widetext}
\begin{equation}\label{mirrev}
M^S(\mathbf{r,r}^{\prime }\mathbf{)=}
\begin{pmatrix}
0&0&0\\
0&\quad
\left( \frac \partial {\partial \mathbf{r}^{\prime }}\frac
\partial {\partial \mathbf{r}}+\mathbf{1}\frac \partial {\partial
\mathbf{r}^{\prime}}\cdot\frac \partial {\partial \mathbf{r}}\right) \eta T\delta +2\frac \partial {\partial \mathbf{r}}\frac \partial {\partial \mathbf{r}
^{\prime }}\widehat{k}T\delta&
\frac \partial {\partial \mathbf{r}}\cdot\eta T\overset{.}{\gamma
}\delta +\frac \partial {\partial \mathbf{r}}\widehat{k}T(\mathrm{tr}[\overset{.}{\gamma }])\delta\\[10pt]
0&
\frac \partial {\partial \mathbf{r}^{\prime }}\cdot\eta T\overset{.}{\gamma }
\delta +\frac \partial {\partial \mathbf{r}^{\prime }}\widehat{k}T(\mathrm{tr}[\overset{.}{\gamma }])\delta &\centering
\frac 12\eta T\overset{.}{\gamma }:\overset{.}{\gamma }\delta
-\frac \partial {\partial \mathbf{r}}\cdot\frac \partial {\partial \mathbf{r}^{\prime}}KT^2\delta +\frac 12\widehat{k}T(\mathrm{tr}[\overset{.}{\gamma }])^2\delta
\end{pmatrix}
\end{equation}
\end{widetext}
\noindent where
\begin{equation}
\overset{.}{\gamma}=\frac{\partial}{\partial\mathbf{r}
}\mathbf{v(r})+\left[  \frac{\partial}{\partial\mathbf{r}}\mathbf{v(r}
)\right]  ^{T},\qquad\widehat{k}=\frac{\zeta}{2}-\frac{\eta}{3}.
\end{equation}
Integrating by parts over $\mathbf{r}^{\prime},$ we get the matrix $M$ which depends only on $\mathbf{r}$
\begin{widetext}
\begin{equation}
M^S(\mathbf{r)=}
\begin{pmatrix}
0 & 0 & 0 \\
0 &\quad
-\left( \frac \partial {\partial \mathbf{r}}\eta T\frac \partial
{\partial \mathbf{r}}+\mathbf{1}\frac \partial {\partial
\mathbf{r}}\cdot\eta T\frac\partial {\partial \mathbf{r}}\right)^T-2\frac \partial {\partial \mathbf{r}}\widehat{k}T\frac \partial
{\partial \mathbf{r}}&
\frac \partial {\partial \mathbf{r}}\cdot\eta T\overset{.}{\gamma
}+\frac\partial{\partial \mathbf{r}}\widehat{k}T(\mathrm{tr}[\overset{.}{\gamma }])\\[10pt]
0 &
-\eta T\overset{.}{\gamma }\frac \partial {\partial \mathbf{r}}-\widehat{
k}T(\mathrm{tr}[\overset{.}{\gamma }])\frac \partial {\partial \mathbf{r}}
&
\frac 12\eta T\overset{.}{\gamma }:\overset{.}{\gamma }-\frac
\partial{\partial \mathbf{r}}\cdot KT^2\frac \partial {\partial \mathbf{r}}+\frac 12\widehat{k}T(\mathrm{tr}[\overset{.}{\gamma }])^2
\end{pmatrix}
\end{equation}
\end{widetext}
The elements of this matrix are differential operators acting on the terms on its right including the vector $\delta S/\delta x$. the matrix  $M^{S}(\mathbf{r,r}^{\prime}\mathbf{)}$ satisfies the required degeneracy condition
\begin{equation}
M\cdot\frac{\delta E}{\delta x}=\int M^{S}(r,r^{\prime})\cdot\left(
\begin{array}
[c]{c}
-\frac{1}{2}\mathbf{v(r)}^{2}\\
\mathbf{v(r)}\\
1
\end{array}
\right) d^{3}r^{\prime}=0\label{degenera1}
\end{equation}
The physical interpretation of this condition has been given in Section II. We now calculate the FDR for the stress tensor by using the last expression for the dissipative operator. A similar procedure may be followed for the heat flux. We omit it in this work for simplicity. The FDR is given by Eq. (\ref{deltacorrelacion}). As the stochastic term associated with the momentum equation is introduced as a divergence, which we denote by $\overleftrightarrow{\sigma}$, we write firstly the FDR as follows
\begin{equation}
\begin{split}
\left\langle \frac{\partial}{\partial r_{\alpha}}\sigma_{r}^{\alpha\beta}\left(  t\right)  \frac{\partial}{\partial r_{\mu}}\sigma_{r^{\prime}}^{\mu\nu}\left(  t^{\prime}\right)  \right\rangle =&\frac{\partial}{\partial
r_{\alpha}}\frac{\partial}{\partial r_{\mu}^{\prime}}\left\langle \sigma
_{r}^{\alpha\beta}\left(  t\right)  \sigma_{r^{\prime}}^{\mu\nu}\left(
t^{\prime}\right)  \right\rangle\\
&=2k_{B}\delta\left(  t-t^{\prime}\right)
\left\langle M_{rr^{\prime}}^{\beta\nu}\right\rangle \label{correlatensor}
\end{split}
\end{equation}
\noindent where the term $M_{rr^{\prime}}^{\beta\nu}$ is determined by the corresponding elements of the dissipation operator matrix Eq. (\ref{mirrev}). We have explicitly
\begin{equation}
\begin{split}
M_{rr^{\prime}}^{\beta\nu}=&\left(  \frac{\partial}{\partial r_{\beta}^{\prime
}}\frac{\partial}{\partial r_{\nu}}+\delta_{\beta\nu}\frac{\partial}{\partial
r_{\gamma}^{\prime}}\frac{\partial}{\partial r_{\gamma}}\right)  \eta\left(
\mathbf{r}\right)  T\left(  \mathbf{r}\right)  \delta\left(  \mathbf{r-r}%
^{\prime}\right)\\
&+2\frac{\partial}{\partial r_{\beta}}\frac{\partial
}{\partial r_{\nu}^{\prime}}\left(  \frac{1}{2}\zeta\left(  \mathbf{r}\right)
-\frac{1}{3}\eta\left(  \mathbf{r}\right)  \right)  T\left(  \mathbf{r}
\right) \delta\left(  \mathbf{r-r}^{\prime}\right) \label{M1}
\end{split}
\end{equation}
\noindent which may be rewritten as
\begin{equation}
\begin{split}
M_{rr^{\prime}}^{\beta\nu}=&\frac{\partial}{\partial r_{\alpha}}\frac{\partial}{\partial r_{\mu}^{\prime}}\{\left(
\delta^{\alpha\nu}\delta^{\beta\mu}+\delta^{\alpha\mu}
\delta^{\beta\nu}\right) \eta\left( \mathbf{r}\right)
T\left( \mathbf{r}\right)\\
&+2\left( \frac{1}{2}\zeta\left( \mathbf{r}
\right) -\frac{1}{3}\eta\left( \mathbf{r}\right) \right) T\left(
\mathbf{r}\right)  \delta_{\alpha\beta}\delta_{\mu\nu}\}\delta\left(
\mathbf{r-r}^{\prime}\right) \label{M2}
\end{split}
\end{equation}
\noindent Therefore the FDR takes the form
\begin{equation}
\begin{split}
\left\langle \sigma_{r}^{\alpha\beta}\left( t\right)  \sigma_{r^{\prime}}^{\mu\nu}\left( t^{\prime}\right) \right\rangle &=k_{B}\{\left(\delta^{\alpha\nu}\delta^{\beta\mu}+\delta^{\alpha\mu}
\delta^{\beta\nu}\right)\left\langle 2\eta\left(\mathbf{r}\right) T\left(\mathbf{r}\right)\right\rangle\\
&+2\left\langle \left(  \zeta\left(  \mathbf{r}\right)
-\frac{2}{3}\eta\left(  \mathbf{r}\right)  \right)  T\left(  \mathbf{r}\right)  \right\rangle\times\\ &\delta_{\alpha\beta}\delta_{\mu\nu}\}\delta\left(\mathbf{r-r}^{\prime}\right)  \delta\left(  t-t^{\prime}\right)
\label{tfdnewton}
\end{split}
\end{equation}
This result coincides with that of Espa{\~n}ol \cite{espanol} obtained from a projection operator technique. It corresponds to a viscous fluid out of equilibrium and the dissipative coefficients $\zeta$ and $\eta$ are dependent on the position. When the dissipative coefficients are constant and the system is in equilibrium the mean value of the temperature is the equilibrium temperature $T_{0}$ and the FDR becomes
\begin{equation}
\begin{split}
\left\langle \sigma_{r}^{\alpha\beta}\left(  t\right)  \sigma_{r^{\prime}}^{\mu\nu}\left(  t^{\prime}\right)  \right\rangle &=k_{B}\{\left(\delta^{\alpha\nu}\delta^{\beta\mu}+\delta^{\alpha\mu}
\delta^{\beta\nu
}\right) 2\eta T_{0}\\
&+2\left(  \zeta-\frac{2}{3}\eta\right)  T_{0}\times\\
&\delta_{\alpha\beta}\delta_{\mu\nu}\}\delta\left(  \mathbf{r-r}^{\prime
}\right)  \delta\left(  t-t^{\prime}\right) \label{tfdnewton2}
\end{split}
\end{equation}
reproducing the Landau-Lifshitz's FDT for the stress tensor \cite{landau1}.
\section{THE MAXWELL FLUID}
In this Section we deal with a viscoelastic fluid and study the fluctuating properties of this kind of fluids. One way to study viscoelastic fluids from a thermodynamic point of view is by assuming that viscosity and elasticity can be considered separately and that the net effect is additive. Under this assumption one arrives to one of the simplest models for viscoelasticity, namely, the Maxwell model. A complete derivation of the Maxwell model in the macroscopic level may be seen for example in \cite{kreuzer81}. One then may tackle the problem of the fluctuations by inserting the Maxwell model in different thermodynamic frameworks. We concentrate here on three of them. On the one hand, LIT in which the state variables of the system are the usual conserved densities of mass, momentum and energy. In this formalism the entropy is considered a function of the conserved densities solely as mentioned in the previous section. On the other hand, we consider extended theories. Different ways of extending the state variables space in order to deal with viscoelastic fluids exist giving rise to different levels of description. A macroscopic level of extended thermodynamics is obtained when one extends the variable space with the dissipative fluxes of the system, the heat flux and the stress tensor in this case. We get one more microscopic level when the same space is extended with structural variables giving an explicit treatment to some internal degrees freedom of the constituents of the fluid. In this level of description one must assure that the dynamic equations of the more microscopic state variables reproduce the Maxwell equation for the stress tensor. In both of these extended versions of irreversible thermodynamics the entropy is considered a functional of the enlarged variables space and this introduces a rather controversial issue in the problem. Thus the system under study is a viscoelastic fluid modelled as a Maxwell fluid when the relaxation times of the fast properties of the system (dissipative fluxes, structural variables) are of the order of the observation time. One would expect that different thermodynamic frameworks lead to the same physical description of the system. Surprisingly, the results diverge. We discuss, in some detail (the final comments in Section V), how to discern the validity of these diverse results by considering the light dispersion properties of the fluid as obtained from the different thermodynamic frameworks investigated.\\
We first examine the LIT case. This problem must be treated in Fourier space so we firstly write the GENERIC\ equation in that space. Consider then the next notation for the time Fourier transform
\begin{equation*}
\widetilde{f}\left( \varpi \right) =\frac{1}{\sqrt{2\pi }}\int_{-\infty
}^{\infty }f\left( t\right) \exp \left( i\varpi t\right) dt
\end{equation*}
Then the GENERIC equation becomes
\begin{equation}
-i\varpi \widetilde{x}\left( \varpi \right) =\widetilde{L}\ast \widetilde{%
\frac{\delta E}{\delta x}}+\widetilde{M}\ast \widetilde{\frac{\delta S}{%
\delta x}}  \label{generic_fourier}
\end{equation}
where the symbol $\ast $ denotes the convolution of the involved Fourier
transformed functions. It is defined as follows
\begin{equation}
f\ast g=\frac{1}{\sqrt{2\pi }}\int_{-\infty }^{\infty }f\left( t^{\prime
}\right) g\left( \varpi -t^{\prime }\right) dt^\prime  \label{convolution}
\end{equation}
Given the dynamic equations of the system the problem is then to find
matrices $\widetilde{L}$ and $\widetilde{M}$ in the Fourier space which
substituted in Eqs. (\ref{generic_fourier}) lead to the dynamic equations of
the system. We now write the linearized equations of the Maxwell fluid in
the Fourier space
\begin{equation}
-i\varpi \widetilde{\rho }_{1}=-\rho _{0}\frac{\partial }{\partial \mathbf{r}%
}\cdot \widetilde{\mathbf{v}}_{1}  \label{fluid1}
\end{equation}
\begin{equation}
-i\varpi \rho _{0}\widetilde{\mathbf{v}}_{1}=-\frac{\partial \widetilde{p}}{%
\partial \mathbf{r}}-\frac{\partial }{\partial \mathbf{r}}\cdot \widetilde{%
\overleftrightarrow{\mathbf{\tau }}}_{1}  \label{fluid2}
\end{equation}
\begin{equation}
-i\varpi C\widetilde{T}_{1}=-\widetilde{p}\ast \frac{\partial }{\partial
\mathbf{r}}\cdot \widetilde{\mathbf{v}}_{1}-\frac{\partial }{\partial
\mathbf{r}}\cdot \left( -\lambda \frac{\partial \widetilde{T}_{1}}{\partial
\mathbf{r}}\right)  \label{fluid3}
\end{equation}
These are the dynamic equations in the variable thermodynamic space of LIT.
They must be supplemented with the time evolution equation of the
Maxwell-Cattaneo kind for the stress tensor
\begin{equation}
\begin{split}
\left( 1-i\varpi \tau _{r}\right) \widetilde{\overleftrightarrow{\mathbf{%
\tau }}}_{1}&=-2\eta \left( \frac{\partial }{\partial \mathbf{r}}\widetilde{%
\mathbf{v}}_{1}\right) ^{S}-\left( \eta _{v}-\frac{2}{3}\eta \right)\times\\
& \left(
\frac{\partial }{\partial \mathbf{r}}\cdot \widetilde{\mathbf{v}}_{1}\right)
\overleftrightarrow{1}  \label{fluid4}
\end{split}
\end{equation}
Eqs. (\ref{fluid1}-\ref{fluid4}) have been linearized around the equilibrium
state. So we have used the following approximated expressions for the dynamic variables
\begin{equation*}
\widetilde{\rho }=\rho _{0}+\widetilde{\rho }_{1}
\end{equation*}
\begin{equation*}
\widetilde{\mathbf{v}}=\widetilde{\mathbf{v}}_{1}
\end{equation*}
\begin{equation*}
\widetilde{T}=T_{0}+\widetilde{T}_{1}
\end{equation*}
\begin{equation*}
\widetilde{\overleftrightarrow{\mathbf{\tau }}}=\widetilde{%
\overleftrightarrow{\mathbf{\tau }}}_{1}
\end{equation*}
By substituting Eq. (\ref{fluid4}) in Eq. (\ref{fluid2}) the momentum density balance equation takes the form
\begin{equation}
\begin{split}
-i\varpi \rho _{0}\widetilde{\mathbf{v}}_{1}&=-\frac{\partial \widetilde{p}}{%
\partial \mathbf{r}}+\frac{\partial }{\partial \mathbf{r}}\cdot \left( \eta
^{D}\left( \frac{\partial }{\partial \mathbf{r}}\widetilde{\mathbf{v}}%
_{1}\right) ^{S}\right.+\\
&\left. 2\overset{\wedge }{\kappa }^{D}\left( \frac{\partial }{%
\partial \mathbf{r}}\cdot \widetilde{\mathbf{v}}_{1}\right)
\overleftrightarrow{1}\right)  \label{fluid2bis}
\end{split}
\end{equation}
where the dispersive coefficients are given by
\begin{equation*}
\eta ^{D}\left( \varpi \right) =\frac{\left( 1+i\varpi \tau _{r}\right) \eta
}{1+\varpi ^{2}\tau _{r}^{2}}
\end{equation*}
\begin{equation*}
\overset{\wedge }{\kappa }^{D}\left( \varpi \right) =\frac{\left( 1+i\varpi
\tau _{r}\right) }{1+\varpi ^{2}\tau _{r}^{2}}\left( \frac{\eta _{v}}{2}-%
\frac{\eta }{3}\right)
\end{equation*}
We now concentrate on the dissipative part of equations (\ref{fluid1}), (\ref
{fluid2bis}) and (\ref{fluid3}) and propose the next matrix $\widetilde{M}$
\begin{widetext}
\begin{equation}
\widetilde{M}=
\begin{pmatrix}
0 & 0 & 0 \\
0 &\left( \frac{\partial }{\partial \mathbf{r}}\frac{\partial }{\partial
\mathbf{r}^{\prime }}+\overleftrightarrow{1}\frac{\partial }{\partial
\mathbf{r}}\cdot \frac{\partial }{\partial \mathbf{r}^{\prime }}\right) \eta
^{D}\left( \varpi \right) T_{0}\delta +2\frac{\partial }{\partial \mathbf{r}}%
\frac{\partial }{\partial \mathbf{r}^{\prime }}\overset{\wedge }{\kappa }%
^{D}\left( \varpi \right) T_{0}\delta
&
\quad 2\frac{\partial }{\partial \mathbf{r}%
}\cdot \eta ^{D}\left( \varpi \right) T_{0}\widetilde{\overleftrightarrow{%
\overset{\cdot }{\gamma }}}\delta +\frac{\partial }{\partial \mathbf{r}}%
\overset{\wedge }{\kappa }^{D}\left( \varpi \right) T_{0}Tr\widetilde{%
\overleftrightarrow{\overset{\cdot }{\gamma }}}\delta \\[10pt]
0 & \frac{\partial }{\partial \mathbf{r}^{\prime }}\cdot \eta ^{D}\left(
\varpi \right) T_{0}\widetilde{\overleftrightarrow{\overset{\cdot }{\gamma }}%
}\delta +\frac{\partial }{\partial \mathbf{r}^{\prime }}\overset{\wedge }{%
\kappa }^{D}\left( \varpi \right) T_{0}Tr\widetilde{\overleftrightarrow{%
\overset{\cdot }{\gamma }}}\delta & \frac{\partial }{\partial \mathbf{r}}%
\cdot \frac{\partial }{\partial \mathbf{r}^{\prime }}\lambda \widetilde{T}%
\delta
\end{pmatrix} \label{matrisM}
\end{equation}
\end{widetext}
It is a direct task to verify that this matrix when applied on the first
order divergence of the LIT entropy
\begin{equation*}
\frac{\delta S}{\delta x}=\left(
\begin{array}{c}
-\frac{\mu _{0}}{T_{0}} \\
\mathbf{0} \\
\frac{1}{T_{0}}
\end{array}
\right)
\end{equation*}
reproduces the dissipative part of Eqs. (\ref{fluid1}), (\ref{fluid2bis})
and (\ref{fluid3}). Now we use the expression Eq. (\ref{correlatensor}) to
find the fluctuation-dissipation theorem for the stochastic part of Eq. (\ref
{fluid2bis}), which is written as a spatial divergence as in the Newtonian
fluid. The central block of Eq. (\ref{matrisM}) determines the form of the
fluctuation-dissipation theorem. The form of the FDT in the Fourier space is
\begin{equation*}
\left\langle \frac{\partial }{\partial r_{\alpha }}\widetilde{\sigma }%
_{r}^{\alpha \beta }\left( \varpi \right) \ast \frac{\partial }{\partial
r_{\mu }^{\prime }}\widetilde{\sigma }_{r^{\prime }}^{\mu \nu }\left( \varpi
^{\prime }\right) \right\rangle =2k_{B}\widetilde{\delta }\left( \varpi
+\varpi ^{\prime }\right) \ast \left\langle \widetilde{M}_{rr^{\prime
}}^{\beta \nu }\right\rangle
\end{equation*}
The resulting FDT is
\begin{multline}
\left\langle \widetilde{\sigma }_{r}^{\alpha \beta }\left( \varpi \right)
\ast \widetilde{\sigma }_{r^{\prime }}^{\mu \nu }\left( \varpi ^{\prime
}\right) \right\rangle =k_{B}\bigg\{ 2\left( \delta ^{\alpha \nu }\delta
^{\beta \mu }+\delta ^{\alpha \mu }\delta ^{\beta \nu }\right)\times\\
\left. \eta^{D}\left( \varpi \right) T_{0}+2\delta ^{\alpha \beta }\delta ^{\mu \nu }%
\overset{\wedge }{\kappa }^{D}\left( \varpi \right) \right\}\times\\
\delta \left(
\mathbf{r-r}^{\prime }\right) \widetilde{\delta }\left( \varpi +\varpi
^{\prime }\right)
\end{multline}
which corresponds to the prescription for the FDT made by Landau \cite
{landau1} in the case of dispersive transport coefficients.\\
We now examine the case of the extension of the variable space with the dissipative fluxes, specifically, with the traceless stress tensor and its trace in such a way that $x=\left( \rho ,v,\varepsilon ,\overset{o}{\tau },\tau \right),$ we define the state of equilibrium as $x_{o}=\left(\rho_{o},0,\varepsilon _{o},0,0,0\right)$. The balance equations, linearized for simplicity, are given by
\begin{gather}
\frac{\partial }{\partial t}\rho _{1}+\frac{\partial }{\partial r_{\alpha }}
(\rho _{o}v_{\alpha })=0  \label{conservacionecua} \\
\frac{\partial }{\partial t}v_{\alpha }=-\frac{\partial }{\partial r_{\alpha
}}p-\frac{\partial }{\partial r_{\beta }}(\tau \mathbf{\delta }_{\alpha
\beta }+\overset{\circ }{\tau }_{\alpha \beta })  \notag \\
\frac{\partial \varepsilon }{\partial t}=-\frac{\partial }{\partial
r_{\alpha }}(\varepsilon _{o}v_{\alpha })-p\frac{\partial }{\partial
r_{\alpha }}v_{\alpha }+\frac{\partial }{\partial r_{\alpha }}\frac{\partial
}{\partial r_{\alpha }}KT_{1}  \notag
\end{gather}
\noindent and the relaxation equations for the extended part of the variable space are
\begin{equation}
\tau _{2}\frac{\partial \overset{\circ }{\tau }_{\mu \nu }}{\partial t}+\overset{\circ }{\tau }_{\mu \nu }=-2\eta X_{\mu \nu \rho \sigma }\frac{\partial }{\partial r_{\sigma }}v_{\rho }  \label{relaja1}
\end{equation}
\begin{equation}
\tau _{o}\frac{\partial \tau }{\partial t}+\tau =-\zeta \frac{\partial }{\partial r_{\mu }}v_{\mu }  \label{relaja2}
\end{equation}
The entropy of the fluid is assumed to be quadratic in the fluxes \cite{Jou96}
\begin{equation}
S(\mathbf{\rho ,\varepsilon ,}\overset{\circ }{\tau }_{\alpha \beta },\tau)=S_{eq}-\frac{v\tau _{o}}{2\zeta T}\tau ^{2}-\frac{v\tau _{2}}{4\eta T}\overset{\circ }{\tau }_{\alpha \beta }:\overset{\circ }{\tau }^{\alpha\beta}  \label{entropiacuadra}
\end{equation}
Calculating the functional derivative of the entropy with respect to the traceless stress tensor and its trace we get
\begin{equation}
\frac{\delta S}{\delta \overset{\circ }{\tau }_{\alpha \beta }}=-\frac{v\tau
_{2}}{2\eta T}\overset{\circ }{\tau }_{\alpha \beta }\approx -\frac{v\tau
_{2}}{2\eta }\overset{\circ }{\tau }_{\alpha \beta }\frac{1}{T_{o}}\left( 1-%
\frac{T_{1}}{T_{o}}\right)   \label{gradiente1}
\end{equation}
\begin{equation}
\frac{\delta S}{\delta \tau }=-\frac{v\tau _{o}}{\zeta T}\tau \approx -\frac{%
v\tau _{o}}{\zeta }\tau \frac{1}{T_{o}}\left( 1-\frac{T_{1}}{T_{o}}\right)
\label{gradiente2}
\end{equation}
\noindent in such a way that matrix $M$ takes the form
\begin{widetext}
\begin{equation}
M(\mathbf{r,r}^{\prime })=\left(
\begin{array}{ccccc}
0 & 0 & 0 & 0 & 0 \\
0 & 0 & 0 & 0 & 0 \\
0 & 0 & \frac{-K}{\rho _{o}AC_{v}}\frac{\partial ^{2}}{\partial r_{\alpha}\partial r_{\beta }^{\prime }}\delta \delta _{\mu \nu }T_{o}^{2} & 0 & 0 \\
0 & 0 & 0 &
\begin{array}{c}
-\frac{4\eta ^{2}T_{o}}{v\tau _{2}^{2}}\left(
1-\frac{T_{1}}{T_{o}}\right)
^{-1}\times \\[7pt]
\left( \overset{\equiv }{X}:\frac{\partial }{\partial \mathbf{r}^{\prime }}
\mathbf{v}\right) \overset{\circ }{\tau }_{\alpha \beta
}^{-1}\delta
\end{array}
& 0 \\
0 & 0 & 0 & 0 &
\begin{array}{c}
-\frac{\zeta ^{2}T_{o}}{v\tau _{o}^{2}}\left(
1-\frac{T_{1}}{T_{o}}\right)
^{-1}\times \\[7pt]
\frac{\partial }{\partial \mathbf{r}^{\prime }}\cdot\mathbf{v}\tau
^{-1}\delta
\end{array}
\end{array}
\right)\label{matrim}
\end{equation}
\end{widetext}
We note that the two last entries of the vector $\delta E/\delta x$ vanish since the total energy does not depend on the traceless stress tensor and its trace. Therefore, the augmented matrix $M$, Eq. (\ref{matrim}), trivially satisfies the degeneracy condition on the gradient of the energy. Eq. (\ref{deltacorrelacion}) does not lead to the right result for the FDR.  See for example \cite{McKane01}, where stochastic process theory was used to obtain the FDR for the stress tensor of a viscoelastic fluid modelled as a Maxwell fluid. One may also prove that the quadratic form of the entropy, Eq. (\ref{entropiacuadra}) does not satisfy the degeneracy condition, Eq. (\ref{degeneracion1}) with $L$ given in \cite{generic} for the viscous fluid, making this description thermodynamically non-consistent.\\
We now describe the viscoelastic fluid within the extended version given in the original GENERIC's papers \cite{generic}.
The variable space is enlarged with the conformation tensor $\mathbf{c}$ given by
\begin{equation}
\mathbf{c}(\mathbf{r)=}\frac{1}{\eta_{p}}\int\mathbf{QQ}\psi(\mathbf{r,Q})
d^{3}Q\label{conformacion}
\end{equation}
\noindent where $\psi(\mathbf{r,Q})$ is the configurational distribution function, $\mathbf{Q}$ is interpreted as a dumbbell configuration vector, and $\eta_{p}$ is the polymer concentration considered here as constant \cite{generic}.
By considering that the complex character of the viscoelastic fluid has only entropic consequences the total energy of the fluid is written as follows
\begin{equation}
E=\int\left[\frac{1}{2}\frac{\mathbf{u(r)}^{2}}{\rho}+\mathbf{\varepsilon
(r)}\right] d^{3}r\notag\label{energiaviscoela}
\end{equation}
While the entropy functional becomes
\begin{equation}
S=\int\mathbf{s}(\rho(\mathbf{r}),\mathbf{\varepsilon}(\mathbf{r}%
))d^{3}r+S_{p}\label{entropiaviscoela}
\end{equation}
This expression includes the polymeric contribution to the entropy in the term  $S_{p}$\ . This term is given by \cite{generic}
\begin{equation}
S_p=\frac{1}{2}\eta_{p}k_{B}\int\left\{  \mathrm{tr}\left[  \mathbf{1-}
c\mathbf{c(r)}\right]  +\ln\left[  \det c\mathbf{c(r)}\right]\right\}
d^{3}r\label{entropolimerica}
\end{equation}
The osmotic pressure tensor is chosen such that the entropy gradient belongs to the null space of the Poisson operator
\begin{equation}
\overset{=}{\mathbf{\Pi}}=T\left(2\mathbf{c\cdot}\frac{\delta\mathbf{S}_{p}
}{\delta\mathbf{c}}+S_p\mathbf{1}\right) \label{presionosmo}
\end{equation}
Substituting Eq. (\ref{entropolimerica}) in Eq. (\ref{presionosmo}) it is obtained
\begin{equation}
\overset{=}{\mathbf{\Pi}}=\eta_{p}k_{B}T(\mathbf{1}-c\mathbf{c)}%
\label{tensorosmotico}
\end{equation}
The matrix $M$, was proposed by \"{O}ttinger and Grmela in \cite{generic}, as follows
\begin{widetext}
\begin{equation}
M=\begin{pmatrix}
0 & 0 & 0 & 0 \\
0 & -\left( \frac{\partial }{\partial \mathbf{r}}\eta T\frac{\partial }{
\partial \mathbf{r}}+\mathbf{1}\frac{\partial }{\partial \mathbf{r}}\cdot\eta T
\frac{\partial }{\partial \mathbf{r}}\right) ^{T} & \frac{\partial }{
\partial \mathbf{r}}\cdot\eta T\overset{.}{\mbox{\boldmath{$\gamma$}} } & 0 \\[10pt]
0 & -\eta T\overset{.}{\mbox{\boldmath{$\gamma$}} }\frac{\partial }{\partial \mathbf{r}} & \frac{
1}{2}\eta T\overset{.}{\mbox{\boldmath{$\gamma$} }}:\overset{.}{\mbox{\boldmath{$\gamma$}} }-\frac{\partial }{
\partial \mathbf{r}}\cdot KT^{2}\frac{\partial }{\partial \mathbf{r}} & 0 \\[10pt]
0 & 0 & 0 & \frac{2}{\eta _{p}k_{B}c\lambda _{H}}\mathbf{c}
\end{pmatrix}.
\end{equation}
\end{widetext}
In this way the form of matrix $M$, together with the matrix $L$ (which is very similar to Eq. (54) of \cite{generic}) reproduces the Maxwell equation for the time evolution of the osmotic pressure tensor \begin{equation}
\left(  1+\lambda_{H}\frac{D}{Dt}\right)  \frac{\overset{=}{\mathbf{\Pi}}
}{\eta_{p}k_{B}T}=-\lambda_{H}\overset{.}{\mbox{\boldmath{$\gamma$}}}\label{maxwellgeneric}%
\end{equation}
\noindent where $D/Dt$ is the material derivative. However this form of $M$ is not suitable to be applied to obtain the FD relation since the lower right term does not have the proper tensor rank. A simple possible modification of $M$ is
\begin{widetext}
\begin{equation}
M=\begin{pmatrix}
0 & 0 & 0 & 0 \\
0 & -\left( \frac{\partial }{\partial \mathbf{r}}\eta T\frac{\partial }{
\partial \mathbf{r}}+\mathbf{1}\frac{\partial }{\partial \mathbf{r}}\cdot\eta T
\frac{\partial }{\partial \mathbf{r}}\right) ^{T} & \frac{\partial }{
\partial \mathbf{r}}\cdot\eta T\overset{.}{\mbox{\boldmath{$\gamma$}}} & 0 \\[10pt]
0 & -\eta T\overset{.}{\mbox{\boldmath{$\gamma$}} }\frac{\partial }{\partial \mathbf{r}} & \frac{
1}{2}\eta T\overset{.}{\mbox{\boldmath{$\gamma$}} }:\overset{.}{\mbox{\boldmath{$\gamma$}}}-\frac{\partial }{
\partial \mathbf{r}}\cdot KT^{2}\frac{\partial }{\partial \mathbf{r}} & 0 \\[10pt]
0 & 0 & 0 & \frac{2}{\eta _{p}k_{B}c\lambda _{H}}\mathbf{1}\mathbf{c:}
\end{pmatrix},
\end{equation}
\end{widetext}
where we have changed $\mathbf{c\cdot}$ by $\mathbf{1}\mathbf{c:}$\ . In this way, the degeneration condition Eq. (\ref{degeneracion2}) is preserved and the Maxwell equation (\ref{maxwellgeneric}) is again obtained.\\
However we make the reader note the fact that this last form of $M$ leads to a FDR which is in discrepancy with that expected for the osmotic component of the stress tensor in the case of a viscoelastic fluid \cite{Mariano01},\cite{McKane01}. Nevertheless it must be mentioned that the obtained FDR is of the $\delta$-correlated type. This fact will be used in our final discussion.
\section{DISCUSSION}
GENERIC \cite{generic} is a formalism which describes the dynamics of complex fluids in terms of two generators, the total energy and the entropy of the system. The thermodynamic consistency of the time evolution equations of the variable space of the system is assured through the degeneracy conditions, Eqs.(\ref{degeneracion1}) and (\ref{degeneracion2}), on the matrices containing the conservative and dissipative dynamics of the system. We remind the reader that the physical meaning of the two mentioned conditions is the second and first laws of thermodynamics respectively. The thermodynamic consistency of any constitutive model is then assured if the two conditions are satisfied by the constituents of the GENERIC form of the dynamics equations of the system. We also remark the key role played by entropy in determining the explicit expressions for the fluctuation-dissipation relations in the fluctuating hydrodynamics obtained from the GENERIC formulation of the dynamics.\\
We have used the fluctuating formulation of GENERIC to find the FDR for a Maxwell fluid in each one of the three thermodynamic frameworks considered. On the one hand in Section IV we have already given the reasons why the EIT version with the variable space enlarged with the dissipative fluxes does not lead to valid results. We remind the reader that the expression for the entropy which leads to the dynamic equations for the Maxwell fluid does not satisfy the degeneracy condition Eq. (\ref{degeneracion1}). In their papers Grmela and \"{O}ttinger also noted that the quadratic form of the entropy for the Maxwell fluid does not satisfy the condition (\ref{degeneracion1}). Specifically, if one would assumes the quadratic form for the polymeric entropy
\begin{equation}
S_p=-\frac{1}{4}\eta_{p}k_{B}\int\left[  \mathbf{1-}c\mathbf{c(r)}\right]:\left[\mathbf{1-}c
\mathbf{c(r)}\right]d^{3}r,
\label{entropica}
\end{equation}
then the gradient of entropy $\left(S=S_{eq}+S_p\right)$ would not be in the null space of the Poisson operator. Thus, the quadratic form of entropy is not a suitable one from the point of view of its thermodynamic consistency.\\
We now concentrate on LIT and EIT version enlarged with the conformation tensor given in Section IV. In order to discern which of these results is valid we resort to a previous work by V\'{a}zquez and L\'{o}pez de Haro \cite{Mariano01}. In that paper the equilibrium Rayleigh-Brillouin spectrum for the Maxwell fluid was computed. This was done by using a procedure by Mountain and by fluctuating hydrodynamics based on two types of FDR. The first of them was of the $\delta$-correlated type and the second one of the exponentially correlated type which is the one prescribed by Landau and Lifshitz for the case where a dispersive dissipation coefficient is present. In the first case the Maxwell fluid was described in the EIT framework and in the second one the fluid was described within LIT. Mountain and fluctuating hydrodynamics calculations rendered the same result only if the Maxwell fluid was described within LIT and the corresponding exponentially correlated FDR was used. From this fact one should conclude that a $\delta$-correlated FDR does not lead to the correct description of the Rayleigh-Brillouin spectrum. This is the case of the FDR in the extended version of the viscoelastic fluid GENERIC theory. At this point we might conclude a number of facts in spite the situation is rather unsatisfactory. Firstly, we remark that the GENERIC scheme allows an effective definition of the entropy of the system by respecting two fundamental physical principles, namely, the first and second laws of thermodynamics. The EIT extended with the dissipative fluxes leading to the quadratic form for entropy is disqualified by its thermodynamic inconsistency. The EIT extended with the conformation tensor will not render the expected Rayleigh-Brillouin spectrum since the Maxwell fluid's FDR within the formalism takes the form of a $\delta$-correlated FDR. Secondly, LIT is consistent from two points of view. On the one hand, it satisfies the prescribed degeneration conditions of GENERIC and on the other, it describes in the correct way the fluctuating properties of the Maxwell fluid leading to the expected Rayleigh-Brillouin spectrum. Thirdly, thus one is forced to conclude that the extended versions of fluctuating irreversible thermodynamics should be revised. We end this paper with the following conclusion. LIT is a proper thermodynamic framework to deal with even situations out from equilibrium for which the extended theories have claimed to require a set of variables in addition to the usual set of conserved densities of LIT to characterize the thermodynamic state.
\section{Acknowledgement}
We acknowledge Prof. M. L{\'o}pez de Haro for suggesting the considered problem in this paper and for fruitful discussions about it. We thank Prof. H.C. {\"O}ttinger for interesting comments.

\end{document}